\documentclass[12pt,epsf]{article}

\usepackage{epsf}
\usepackage[dvips]{graphicx}

\begin{document}

\begin{center}
{\bfseries SPIN MATRIX ELEMENTS FOR THE ELASTIC PROTON-PROTON AND
PROTON-ANTIPROTON COLLISIONS}

\vskip 5mm

V.A. Okorokov$^{1 \dag}$ and S.B. Nurushev$^{2}$

\vskip 5mm

{\small (1) {\it Moscow Engineering Physics Institute (State
University), \\ Kashirskoe Shosse 31, 115409 Moscow, Russian
Federation }
\\
(2) {\it Institute for High Energy Physics,\\
142284 Protvino, Moscow Region, Russian Federation }
\\
$\dag$ {\it E-mail: Okorokov@bnl.gov; VAOkorokov@mephi.ru}}
\end{center}

\vskip 5mm

\begin{center}
\begin{minipage}{150mm}
\centerline{\bf Abstract} Motivated by the present extensive RHIC
(BNL) spin program and further PAX project at FAIR, we study the
possibility of explicit reconstruction of full set of helicity
amplitudes by joint consideration of elastic proton-proton and
proton-antiproton scattering. Procedure is based on the derivative
relations for the helicity amplitudes, explicit parametrization of
the leading spin non-flip amplitudes and crossing - symmetry
relations. Some preliminary results are presented for the PAX
energies to show the expected magnitudes of the spin dependent
observables as a functions of invariant Mandelstam's variables $s$
and $t$.
\end{minipage}
\end{center}

\vskip 10mm

\section{Introduction}
The quantitative description of high-energy soft processes is not
yet achieved in frame of the QCD. Elastic scattering of hadrons
has always been a crucial tool for study the dynamics of strong
interaction. Despite of important results at the interface between
soft and hard physics and recent progress in nonperturbative QCD,
elastic scattering cannot be described in a pure QCD framework
because of this reaction is soft one essentially. In such a
situation the direct reconstruction of the scattering matrix from
the complete set of the experimental data will be the appropriate
method. One may hope that such spin programs will be realized at
RHIC, FAIR (PAX project) and other facilities in near future.

In order to make the predictions for the observables in those
facilities one needs to have a well justified method, contains a
small number of free parameters and applicable at wide range of
the kinematics variables. Below we propose such a technique. Its
correctness may be tested by the joint consideration of
proton-proton and proton-antiproton elastic scattering data. Thus
the further study of the $pp$ and $p\bar{p}$ elastic scattering in
wide range of momentum transfer and initial energies is extremely
important task.

\section{Preliminary results}
Two methods have been suggested for reconstruction  the full set
of helicity amplitudes for elastic $p\bar{p}$-collisions in
\cite{Okorokov-2005, Okorokov-2006}. We describe in details the
method for deducing the  helicity amplitudes based on
crossing-symmetry relation and the derivative relations. As known
the amplitudes for the binary reaction $A+B \to C+D$ in
$s-,~t-,~\mbox{and}~u-$ channels are described by just one set of
analytic functions. The following preliminary expression for full
set of helicity amplitudes of $p\bar{p}$ elastic scattering via
set of helicity amplitudes for $pp$ elastic reaction has been
derived:

\begin{equation}
\left. \begin{array}{l} \vspace{0.2cm}
 \Phi _1^{p\bar p}  = {1 \mathord{\left/
 {\vphantom {1 2}} \right.
 \kern-\nulldelimiterspace} 2}\,\left[ {\sin ^2 \psi \left( {\Phi _1^{pp}
+ \Phi _2^{pp}  + \Phi _3^{pp} } \right)\, + \left( {1 + \cos ^2
\psi } \right)\Phi _4^{pp} } \right]
 \\
\vspace{0.2cm}
 \Phi _2^{p\bar p}  = {1 \mathord{\left/
 {\vphantom {1 2}} \right.
 \kern-\nulldelimiterspace} 2}\,\left[ {\sin ^2 \psi \left( {\Phi _1^{pp}
+ \Phi _3^{pp}  - \Phi _4^{pp} } \right) - \left( {1 + \cos ^2
\psi }
\right)\Phi _2^{pp} } \right]\, \\
\vspace{0.2cm}
 \Phi _3^{p\bar p}  = {1 \mathord{\left/
 {\vphantom {1 2}} \right.
 \kern-\nulldelimiterspace} 2}\,\left[ {\sin ^2 \psi \left( {\Phi _1^{pp}
+ \Phi _2^{pp}  - \Phi _4^{pp} } \right) - \left( {1 + \cos ^2
\psi }
\right)\Phi _3^{pp} } \right] \\
\vspace{0.2cm}
 \Phi _4^{p\bar p}  = {1 \mathord{\left/
 {\vphantom {1 2}} \right.
 \kern-\nulldelimiterspace} 2}\,\left[ {\left( {1 + \cos ^2 \psi }
\right)\Phi _1^{pp}  - \sin ^2 \psi \left( {\Phi _3^{pp}  + \Phi
_2^{pp}  -
\Phi _4^{pp} } \right)} \right] + 2\Phi _5^{pp} \sin \psi  \\
\vspace{0.2cm}
 \Phi _5^{p\bar p}  = {1 \mathord{\left/
 {\vphantom {1 2}} \right.
 \kern-\nulldelimiterspace} 2}\cos \psi \left[ {\sin \psi \left( {\Phi
_1^{pp}  + \Phi _2^{pp}  + \Phi _3^{pp}  - \Phi _4^{pp} } \right)
+ 2\Phi
_5^{pp} } \right] \\
 \end{array} \right\}\label{eq:helamp-pantip-sys}
\end{equation}
where
$$\cos \psi  =  \sqrt {\frac{\textstyle st}{\textstyle \left(s -
4m_{p}^{2} \right)\left(t - 4m_{p}^{2}\right)}};~\sin \psi  =
2m_{p} \sqrt{\frac{\textstyle 4m_{p}^{2}-s-t}{\textstyle \left(s -
4m_{p}^{2}\right)\left(t - 4m_{p}^{2}\right)}};$$ $\psi$ - the $t
\to s$ crossing angle \cite{Leader-2001}, $m_{p}$ - proton mass.
The system (\ref{eq:helamp-pantip-sys}) shows apparent analytical
forms for the full set of amplitudes of elastic $p\bar{p}$
scattering $\left\{ {\Phi _{i}^{p\bar p} } \right\}_{i = 1-5}$ via
helicity amplitudes $\left\{ {\Phi _{j}^{pp} } \right\}_{j = 1-5}$
for $pp$ channel. It should be noted that the relation $\Phi
_{5}^{pp}=-\Phi _{6}^{pp}$ has been taken into account in the
(\ref{eq:helamp-pantip-sys}) already.

One must consider $pp$ elastic scattering in order to obtain some
additional relations for set of helicity amplitudes $\left\{ {\Phi
_{j}^{pp} } \right\}_{j = 1-5}$. According to the model
independent analysis of unpolarized differential cross section
\cite{Phillips-1973} the following parametrization is suggested
for spin non-flip amplitude $\Phi_{1}^{pp}$:
\begin{equation}
\Phi_{1}^{pp}=i\sqrt{A}\exp\left[\left( \frac{\textstyle
B}{\textstyle 2}+\alpha'ln\left(\frac{\textstyle s}{\textstyle
s_{0}}\right)-i\alpha'\frac{\textstyle \pi}{\textstyle
2}\right)t\right] + \left( \frac{\textstyle
\sqrt{C_{0}}}{\textstyle s}-i\sqrt{C_{\infty}}\right)\exp(Dt/2)
\label{eq:pp-Phil-param},
\end{equation}
where $s_{0}=1\mbox{GeV}^{2}$ and $\sqrt{A}$, $B$, $\alpha'$,
$\sqrt{C_{0}}$, $\sqrt{C_{\infty}}$, $D$ are free parameters
obtained from experimental data for elastic $pp$ reaction in the
interval $0.15 \! \leq \mid \! t \! \mid \leq \! 5.0~
\mbox{GeV}^{2}$ \cite{Phillips-1973}.

The two different assumptions were made in calculations: 1 -
$\Phi_{2}^{pp} = 0$ and $\Phi_{1}^{pp} = \Phi_{3}^{pp}$; 2 -
$\Phi_{2}^{pp} = -\Phi_{4}^{pp}$ and $\Phi_{1}^{pp} =
\Phi_{3}^{pp}$. Similar assumptions were used in the analysis of
$pp$ elastic collisions at RHIC \cite{Okorokov-2001} and for
recent study of $p\bar{p}$ elastic interactions in PAX
\cite{Okorokov-2005} by another method of reconstruction the full
set of helicity amplitudes for nucleon-nucleon elastic scattering.

 The derivative relations allow to express the
spin single-flip amplitude $\Phi_{5}^{pp}$ and the spin
double-flip amplitude $\Phi_{4}^{pp}$ via $\Phi_{1}^{pp}$
\cite{Schrempp-1975}:
\begin{equation}
\Phi_{5}^{pp}\left(s,t\right)=C_{1}^{pp}(s)\frac{\textstyle
\partial}{\textstyle \partial(\sqrt{-t})}\Phi_{1}^{pp}\left(s,t\right);~~
\Phi_{4}^{pp}\left(s,t\right)=C_{2}^{pp}(s)\frac{\textstyle
\partial^{2}}{\textstyle
\partial(\sqrt{-t})^{2}}\Phi_{1}^{pp}\left(s,t\right),
\label{derriv-rel-my}
\end{equation}
where $C_{k}^{pp}(s)=C_{k1}^{pp}(s)+iC_{k2}^{pp}(s),~k=1,2$ -
complex parameters in general.

The Odderon hypothesis is important for definition of unknown
parameters $C_{k}(s)^{pp}$ in the derivative relations
(\ref{derriv-rel-my}). We use the asymptotic total cross section,
differential cross section, $\rho$ and $B$ parameters both in $pp$
and $p\bar{p}$ elastic reactions in order to obtain the complex
unknown parameters. The suggested reconstruction procedure is
described in details for asymptotic energies in
\cite{Okorokov-2006}. According to \cite{Schrempp-1975} one can
asume that $C_{k}(s)$ are real constants at low and medium
energies for $pp$ scattering. Numerical values of these parameters
have been found in crude approach from comparison with early
results for $pp$ differential cross section \cite{Rarita-1968,
Austin-1970} at qualitatively level: $C_{1}=-0.05~\mbox{GeV}$ and
$C_{2}=0.02~\mbox{GeV}$. But further study is planed for this
hypothesis.

We present below preliminary results for elastic $p\bar{p}$
collisions at energy range of PAX experiment obtained by discussed
above reconstruction procedure. We have calculated
$|t|$-dependence of differential cross-section, polarization, some
elements of the second-rank spin tensors, namely, spin-correlation
$\left(C_{ik}\right)$ and depolarization $\left(D_{ik}\right)$
tensors and Wolfenstein parameters. These parameters have been
calculated according to the general analytical formulas in
Regge-pole theory \cite{Leader-1966} for 1-st and 2-d versions of
assumptions for the $pp$ helicity amplitudes described above.

The preliminary $t$-dependence is shown for each of observable in
fig.1 and in fig.2 for 1-st and 2-d version of assumptions for a
range of $0.15 \! \leq \mid \! t \! \mid \leq \! 5.0~
\mbox{GeV}^{2}$. The validity region for (\ref{derriv-rel-my}) is
$\left(-t\right) \! \ll \! s$ \cite{Schrempp-1975} but the
derivative relations are used at all values of $t$-invariant
allowed by parameterization (\ref{eq:pp-Phil-param}) even at
lowest energy $\sqrt{S}=3$ GeV. One can see that all dependences
are smooth at this energy and no any dramatic features at large
$|t|$ are seen. But the applicability of derivative relations at
large $|t|$ requires additional study.

One can see that differential cross section is almost the same for
two versions in the domain of $|t| \! \geq \! 2.0~\mbox{GeV}^{2}$
but there is difference between $d\sigma/dt$ values at smaller
$|t|$ for various versions (fig.1a, 2a) and the value of this
difference increases with energy increasing. The maximum value of
this difference is at $|t| \! \simeq \! 1.0 - 1.5~\mbox{GeV}^{2}$
and removes to larger $|t|$ at energy increasing. There is a
qualitative agreement between our predictions and the experimental
data \cite{Rarita-1968,Austin-1970,Bruneton-1975}. Polarization
(fig.1b, 2b) shows good agreement with early Regge model
predictions and three-pole fits \cite{Austin-1970} at lower
energies but our calculation predicts opposite sign and different
magnitude of polarization for $|t| \! \leq \! 1.0~\mbox{GeV}^{2}$
at highest energy $\sqrt{S}=14.7$ GeV. One needs more precise
experimental data for separating unambiguously between two
versions of our calculations (fig.1b, 2b). Thus we have obtained
correct energy dependence of polarization at qualitative level. It
is worthwhile to note, that the sign on polarization for
$p\bar{p}$ scattering is opposite to the sign of $P$ for elastic
$pp$ scattering independent of versions of our calculations.

The depolarization tensor depends on $|t|$ in similar way for both
versions at large values $|t| \geq 1.5~\mbox{GeV}^{2}$ (fig.1c,
2c). But values of $D$ are different for 1-st and 2-d version of
assumption significantly at low $|t|$. The element $C_{NN}$ of
spin-correlation tensor depends on initial energy at $|t| \leq
2.0~\mbox{GeV}^{2}$ (fig.1d) but $C_{NN}$ is almost independent of
$s$ and $|t|$ at larger $|t|$ (fig.1d, 2d). Our preliminary
results differ from early predictions \cite{Rarita-1968} at all
energies under study. The element of spin-correlation tensor
$C_{kp}$ shows non trivial $|t|$-dependence (fig.1e, 2e).

The $|t|$-dependences for the first pair of Wolfenshtein
parameters $R$ and $A$ are shown at fig.1f (fig.2f) and fig.1g
(fig.2g) for 1-st (2-d) version assumptions for $pp$ helicity
amplitudes respectively. The results for first ($A$ and $R$) pair
of Wolfenstein parameters differ from predictions based on
Regge-pole model significantly \cite{Rarita-1968}. But it should
be emphasized that the early predictions were made for more narrow
range $|t| \! \leq \! 1.0~\mbox{GeV}^{2}$. The preliminary results
for second pair of Wolfenshtein parameters $R'$ and $A'$ are shown
in fig.1h (fig.2h) and fig.1i (fig.2i) respectively. All
Wolfenshtein parameters show the $|t|$-dependences which are
similar each to other at qualitative level for all energies with
the exception for the lowest one $\sqrt{S}=3$ GeV.

\section{Conclusion}

The main results of this paper are following. We propose the
analytical method for calculation of the full set of helicity
amplitudes for elastic $p\bar{p}$ reaction. This method is based
on fundamental crossing-symmetry property and derivative relations
for helicity amplitudes. Using this method the preliminary
expressions for $p\bar{p}$ helicity amplitudes were derived. The
known approximation of $pp$ helicity amplitudes (at least
$\Phi_{1}^{pp}$) and some supplementary conditions are used in
order to obtain the full set of helicity amplitudes for elastic
$p\bar{p}$ scattering in wide range of $|t|$ parameter for the
first time. The problem of definition of free parameters from the
derivative relations is discussed. The preliminary numerical
results are obtained for wide a set of observables in intermediate
(PAX) energy domain. Our results agree with Regge-pole model
predictions for some parameters qualitatively. The unified
analysis of the $pp$ and $p\bar{p}$ data and new experimental
results will allow to check in details the proposed method. It
seems this analytical procedure might be useful for direct
reconstruction of the spin matrix elements in elastic $pp$ and
$p\bar{p}$ scattering at wide initial energy and square of
momentum transfer domains.

\begin{figure}[htb]
\parbox[t]{5.1cm}{%
\epsfysize=51mm \centerline{
\epsfbox{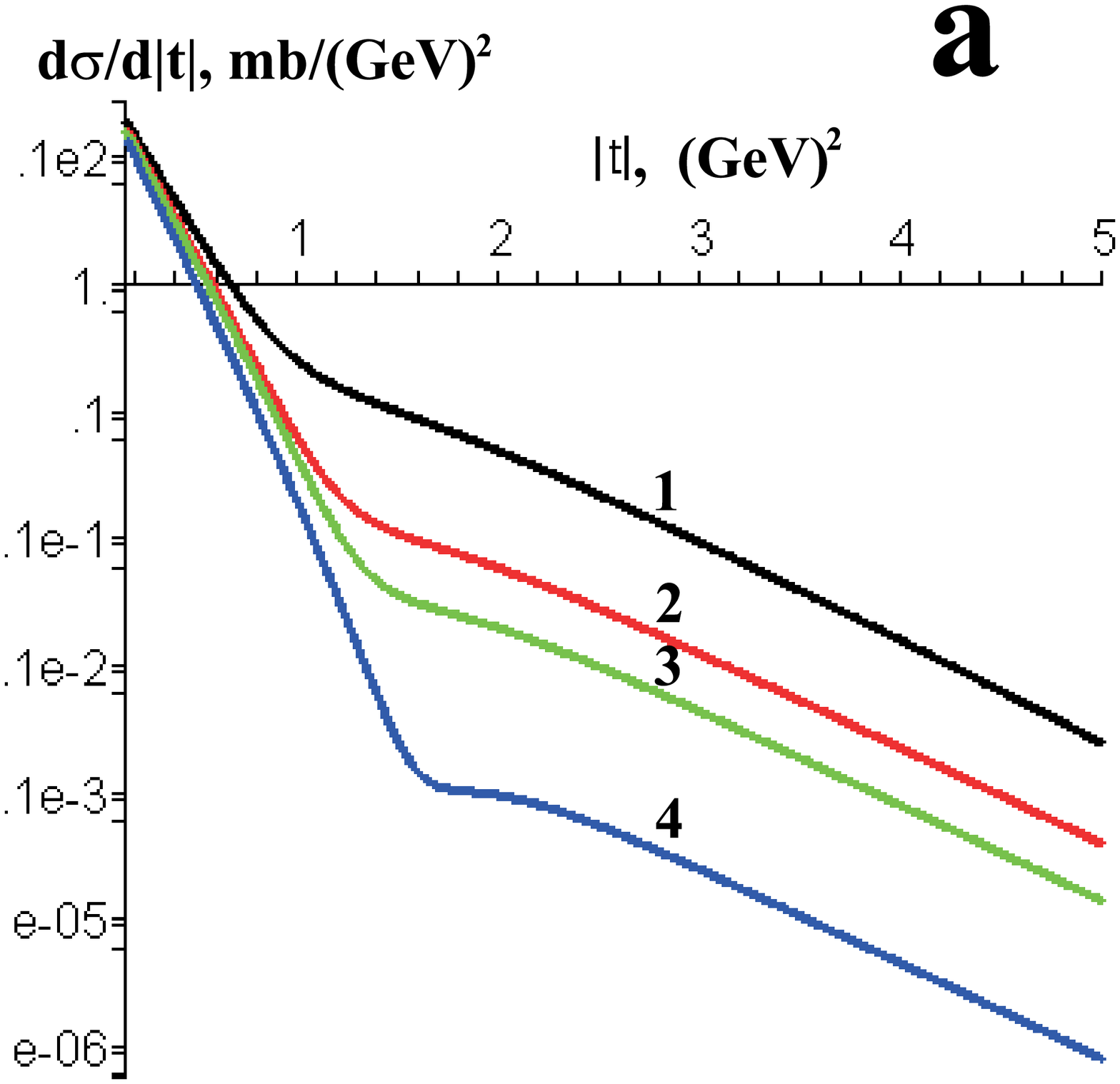}
}\label{ppbar-PAXenergies-dSigma-version-1}}\quad
\parbox[t]{5.1cm}{%
\epsfysize=51mm \centerline{
\epsfbox{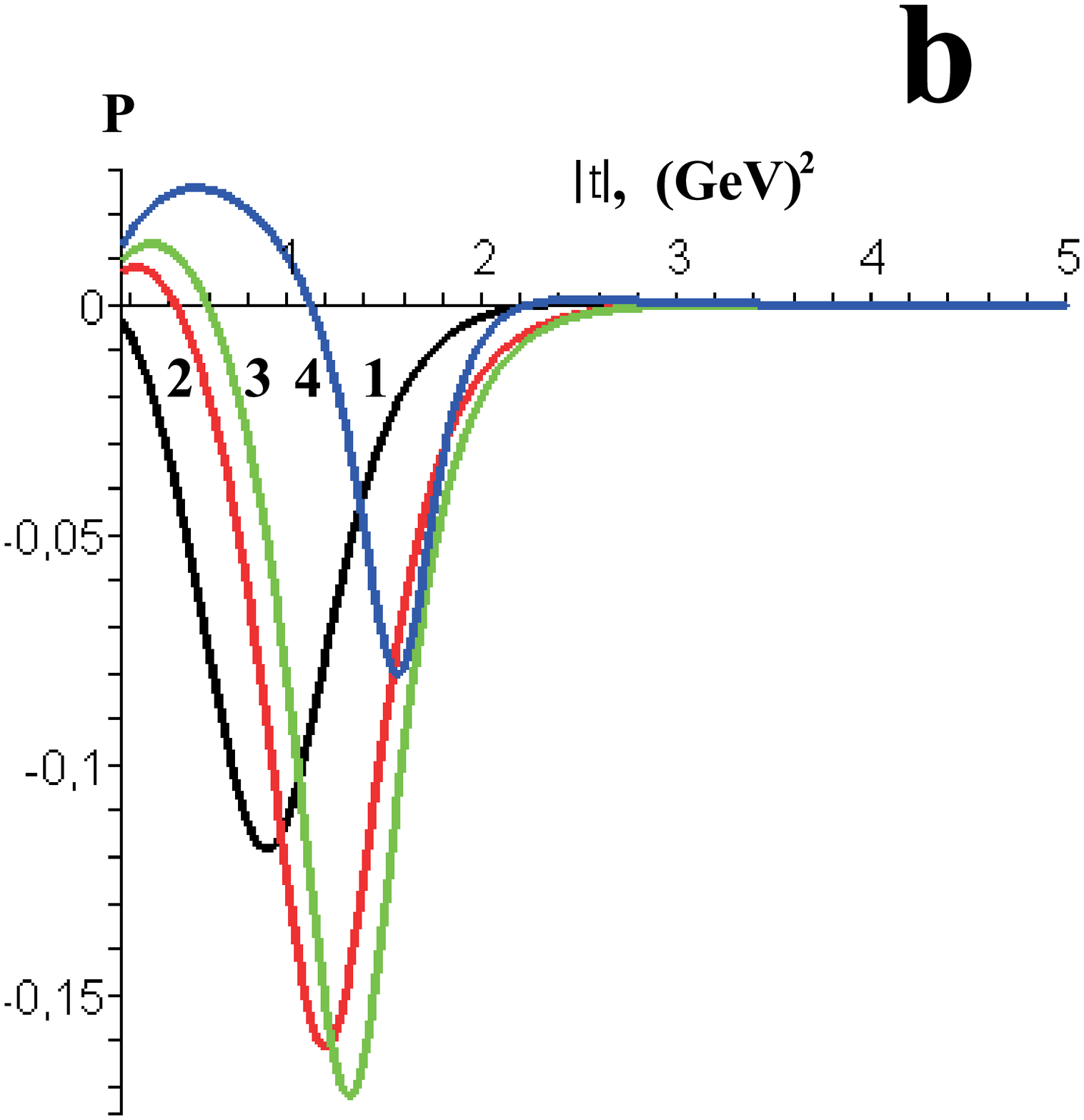}
}\label{ppbar-PAXenergies-P-version-1}}\quad
\parbox[t]{5.1cm}{%
\epsfysize=51mm \centerline{
\epsfbox{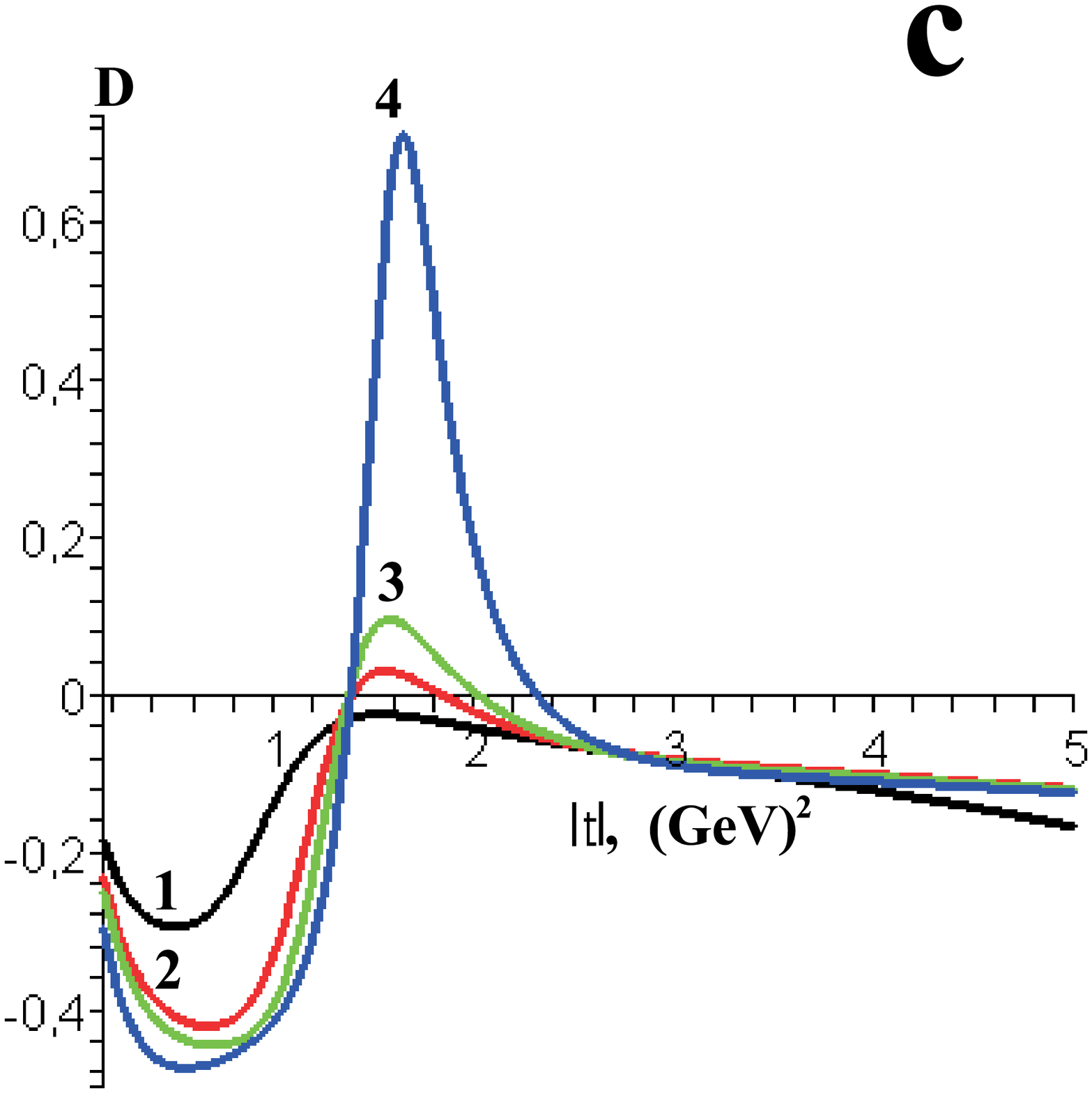}
}\label{ppbar-PAXenergies-D-version-1}}\\\\
\parbox[t]{5.1cm}{%
\epsfysize=51mm \centerline{
\epsfbox{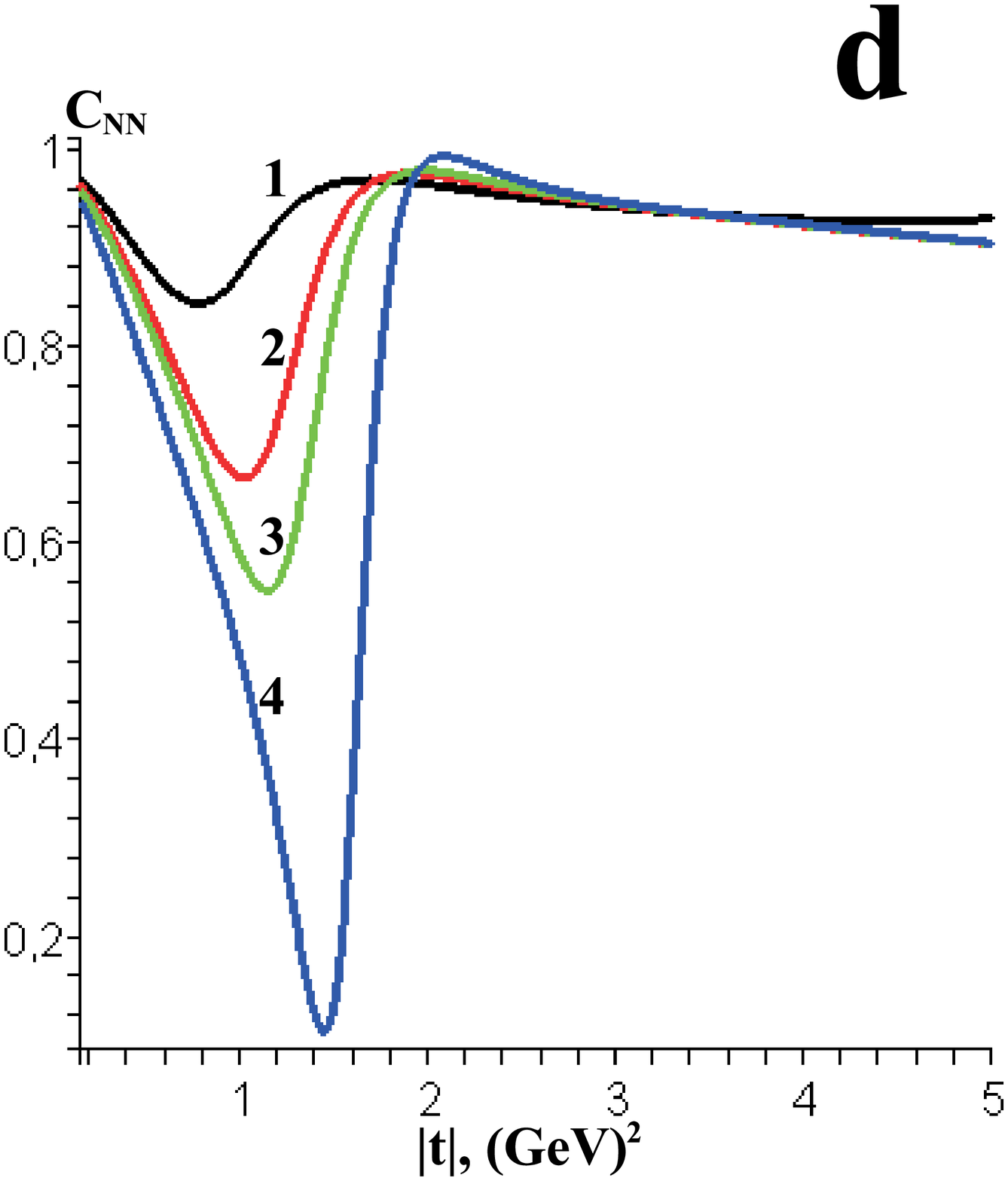}
}\label{ppbar-PAXenergies-CNN-version-1}}\quad
\parbox[t]{5.1cm}{%
\epsfysize=51mm \centerline{
\epsfbox{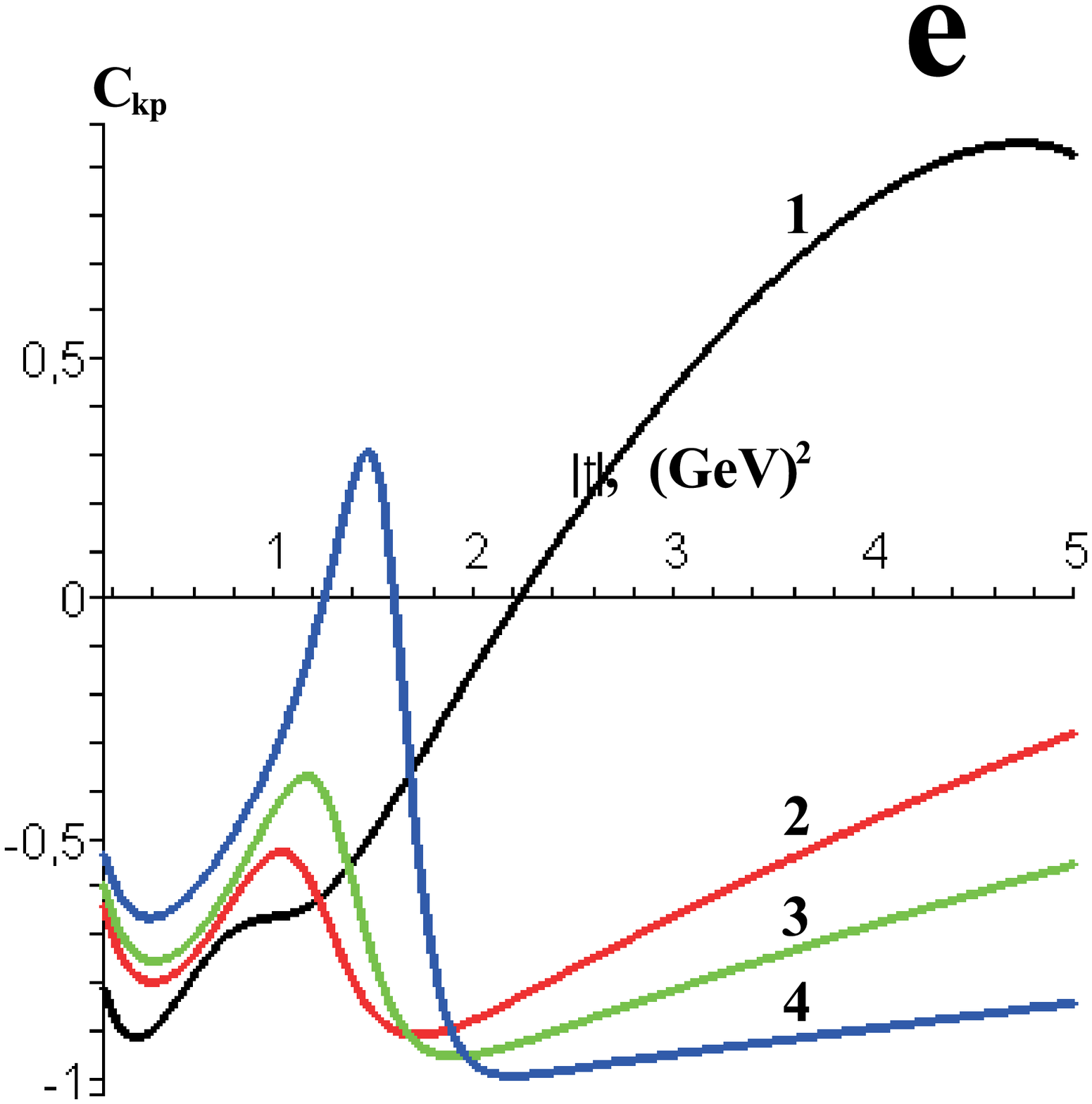}
}\label{ppbar-PAXenergies-Ckp-version-1}}\quad
\parbox[t]{5.1cm}{%
\epsfysize=51mm \centerline{
\epsfbox{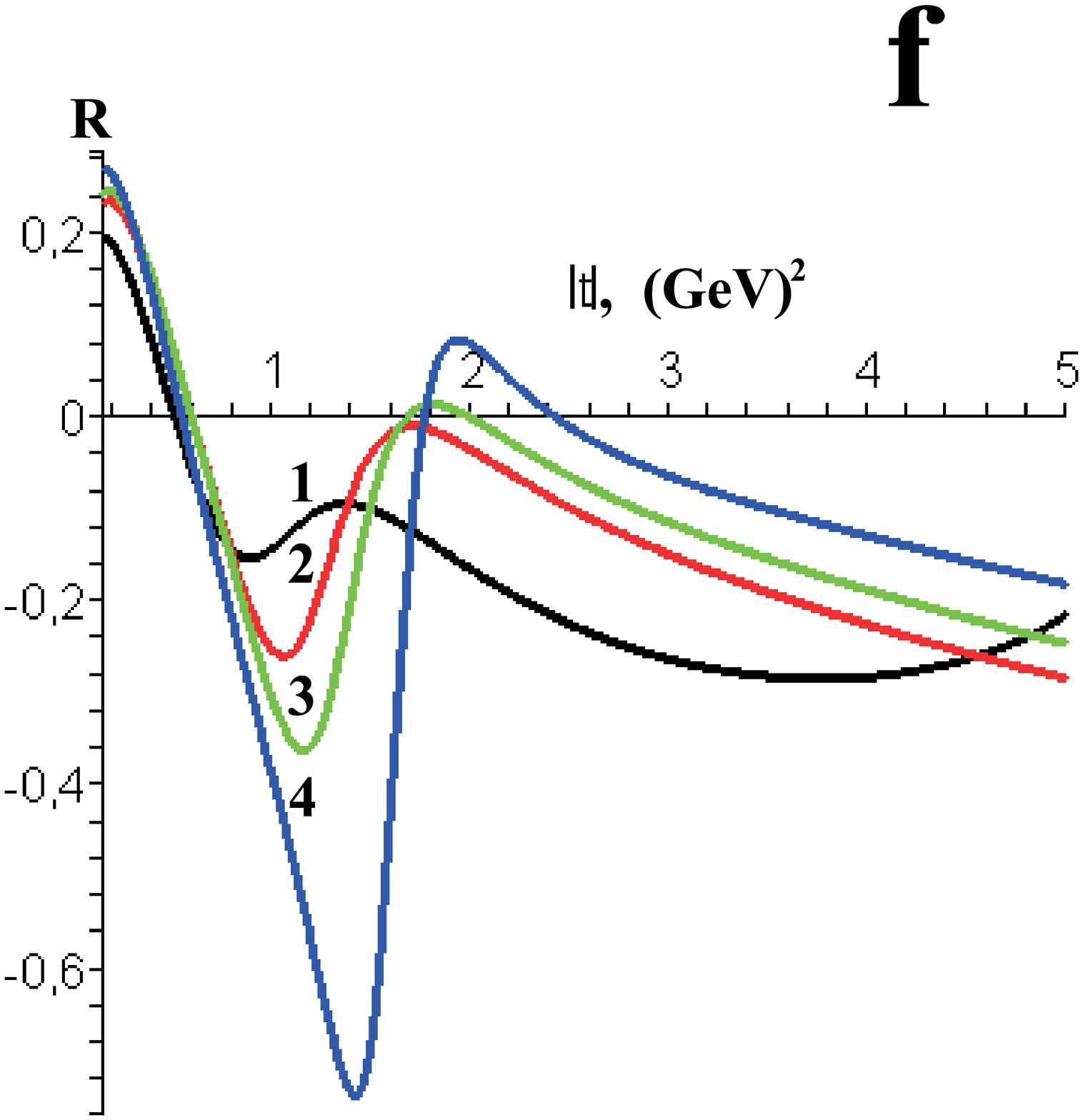}
}\label{ppbar-PAXenergies-R-version-1}}\\\\
\parbox[t]{5.1cm}{%
\epsfysize=51mm \centerline{
\epsfbox{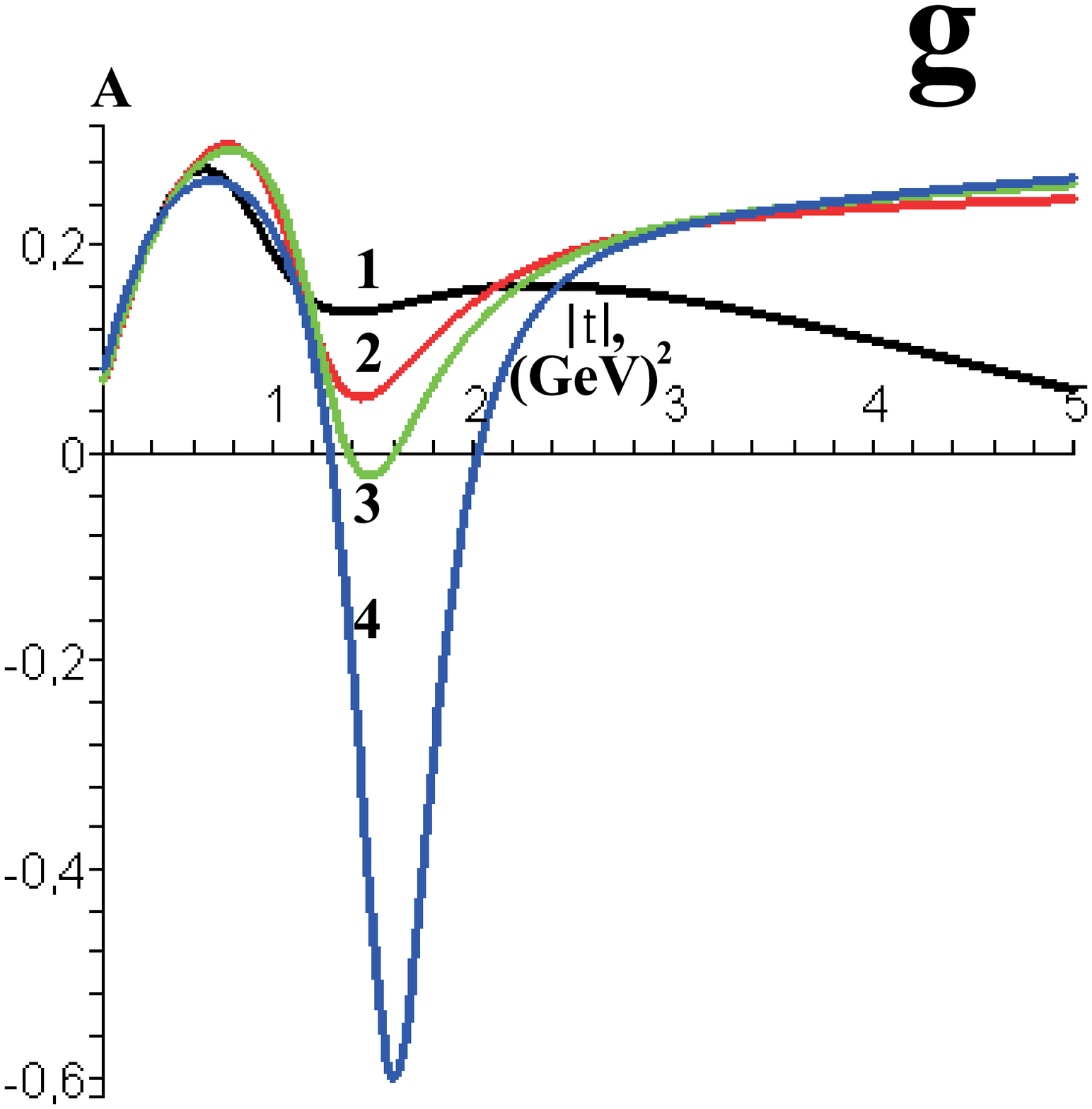}
}\label{ppbar-PAXenergies-A-version-1}}\quad
\parbox[t]{5.1cm}{%
\epsfysize=51mm \centerline{
\epsfbox{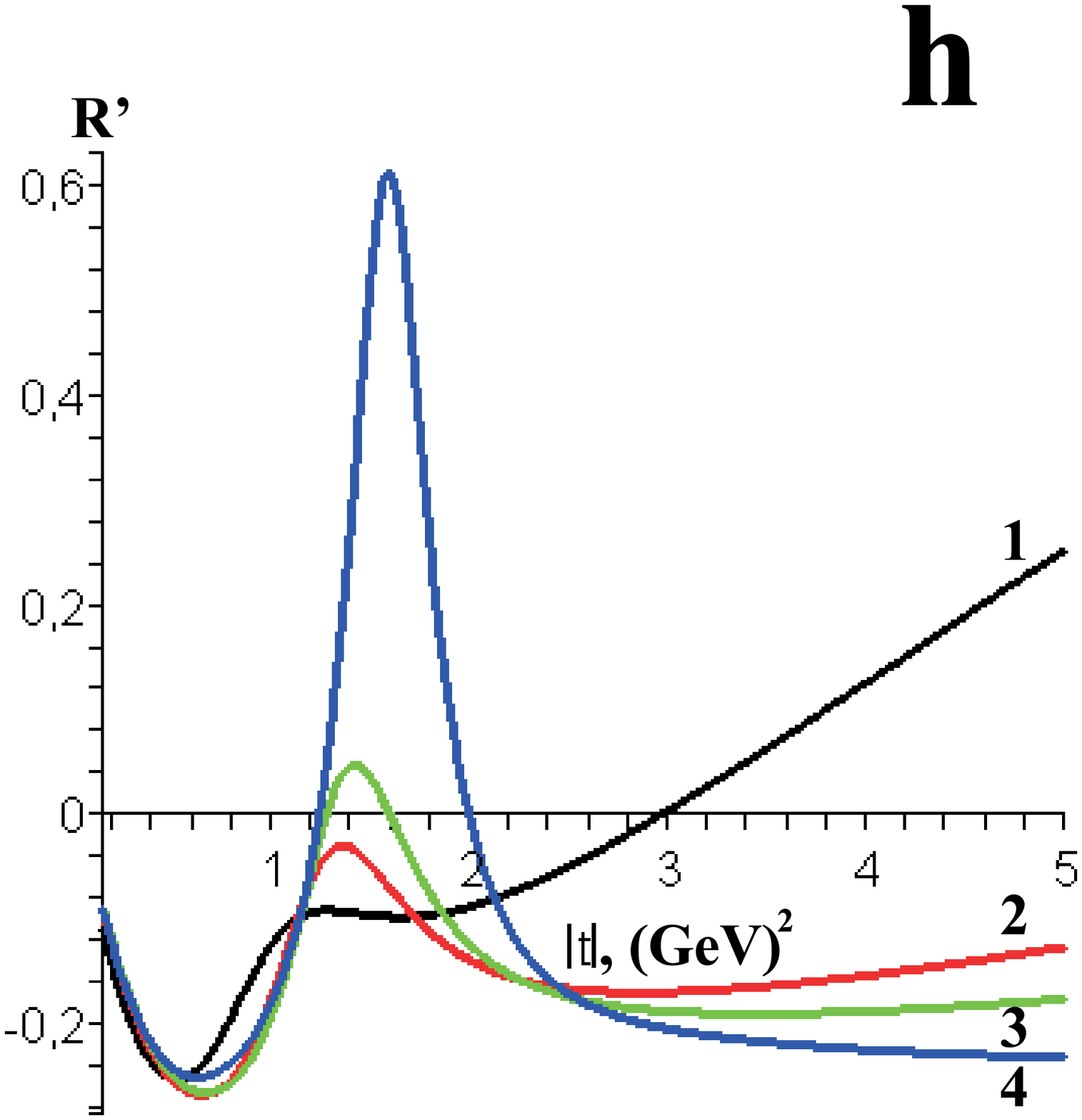}
}\label{ppbar-PAXenergies-Rprim-version-1}}\quad
\parbox[t]{5.1cm}{%
\epsfysize=51mm \centerline{
\epsfbox{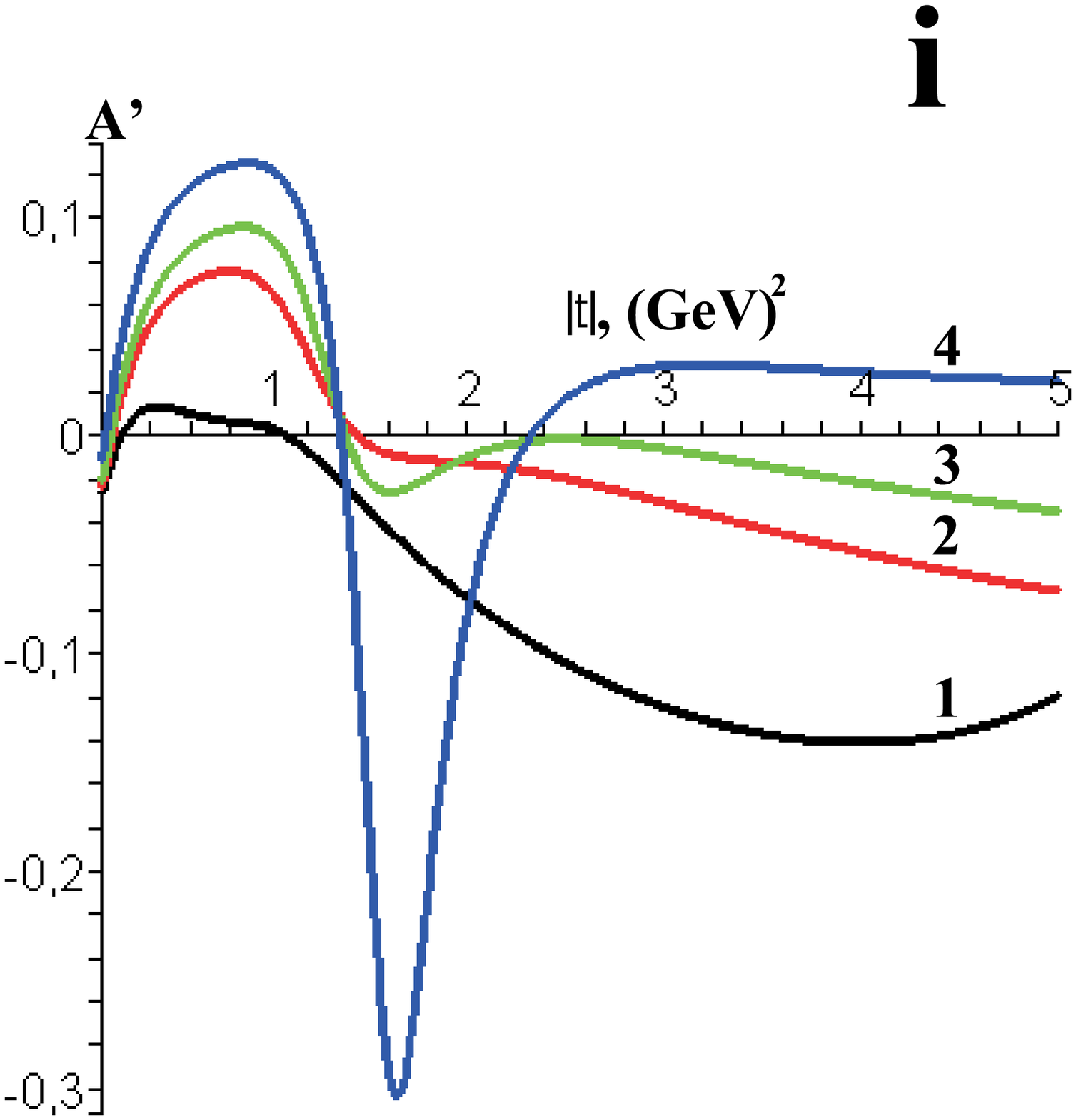}
}\label{ppbar-PAXenergies-Aprim-version-1}}
\caption{$t$-Dependence of some observable parameters for elastic
$p\bar{p}$ scattering in intermediate (PAX) energy domain: 1 -
$\sqrt{S}=3$ GeV (black), 2 - $\sqrt{S}=5$ GeV (red), 3 -
$\sqrt{S}=6.5$ GeV (green), 4 - $\sqrt{S}=14.7$ GeV (blue). The
first version of additional relations between $pp$ helicity
amplitudes is under consideration: \emph{a} - differential cross
section, \emph{b} - polarization, \emph{c} - depolarization
parameter, \emph{d} - correlation of the normal components of
polarization, \emph{e} - correlation of the transverse components
of polarization, \emph{f} - transverse polarization rotation
parameter, \emph{g} - longitudinal polarization rotation
parameter, \emph{h} - correlation of transverse-longitudinal
components of polarization, \emph{i} - correlation of
longitudinal-longitudinal components of polarization.}
\end{figure}

\begin{figure}[htb]
\parbox[t]{5.1cm}{%
\epsfysize=51mm \centerline{
\epsfbox{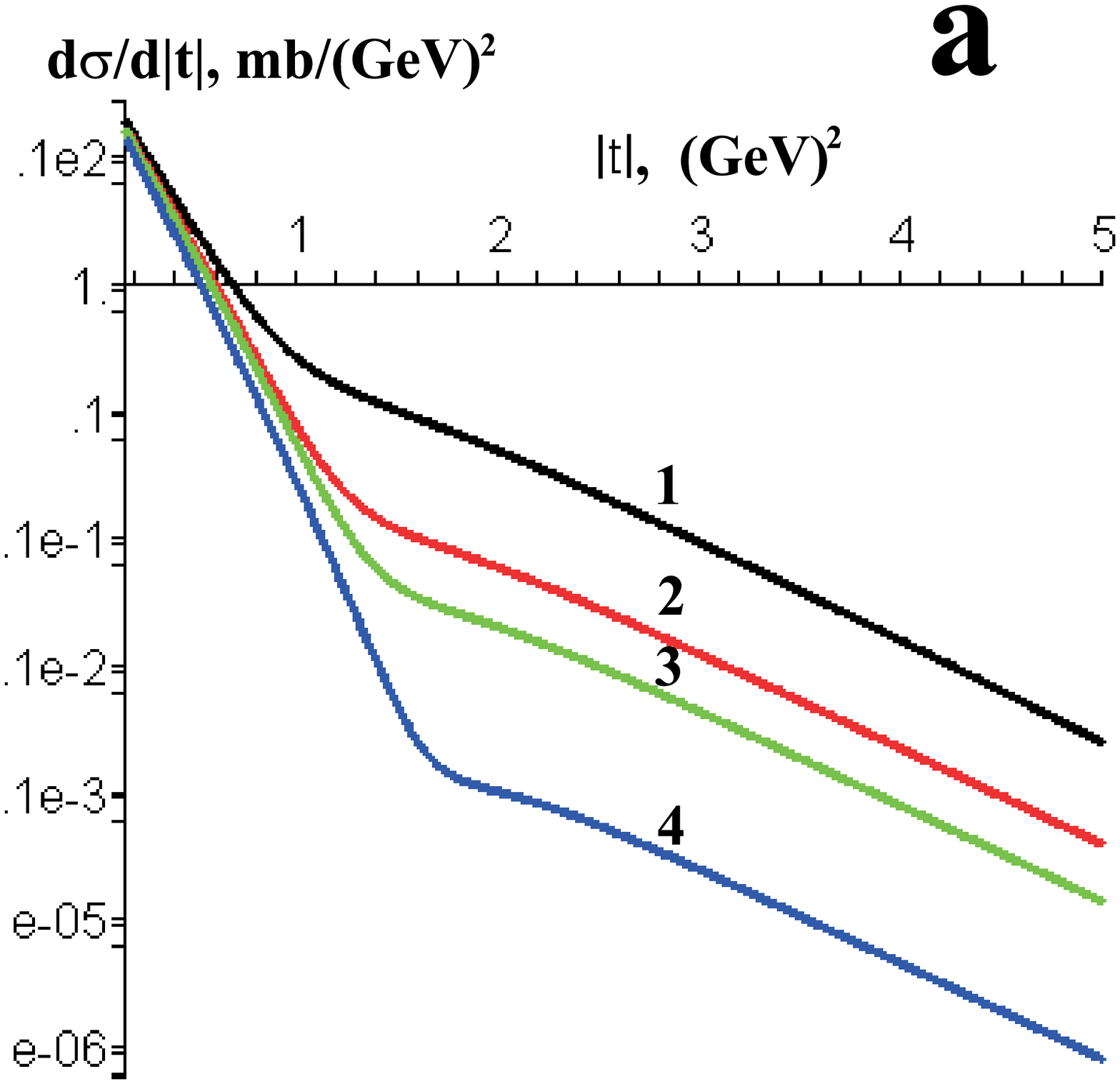}
}\label{ppbar-PAXenergies-dSigma-version-2}}\quad
\parbox[t]{5.1cm}{%
\epsfysize=51mm \centerline{
\epsfbox{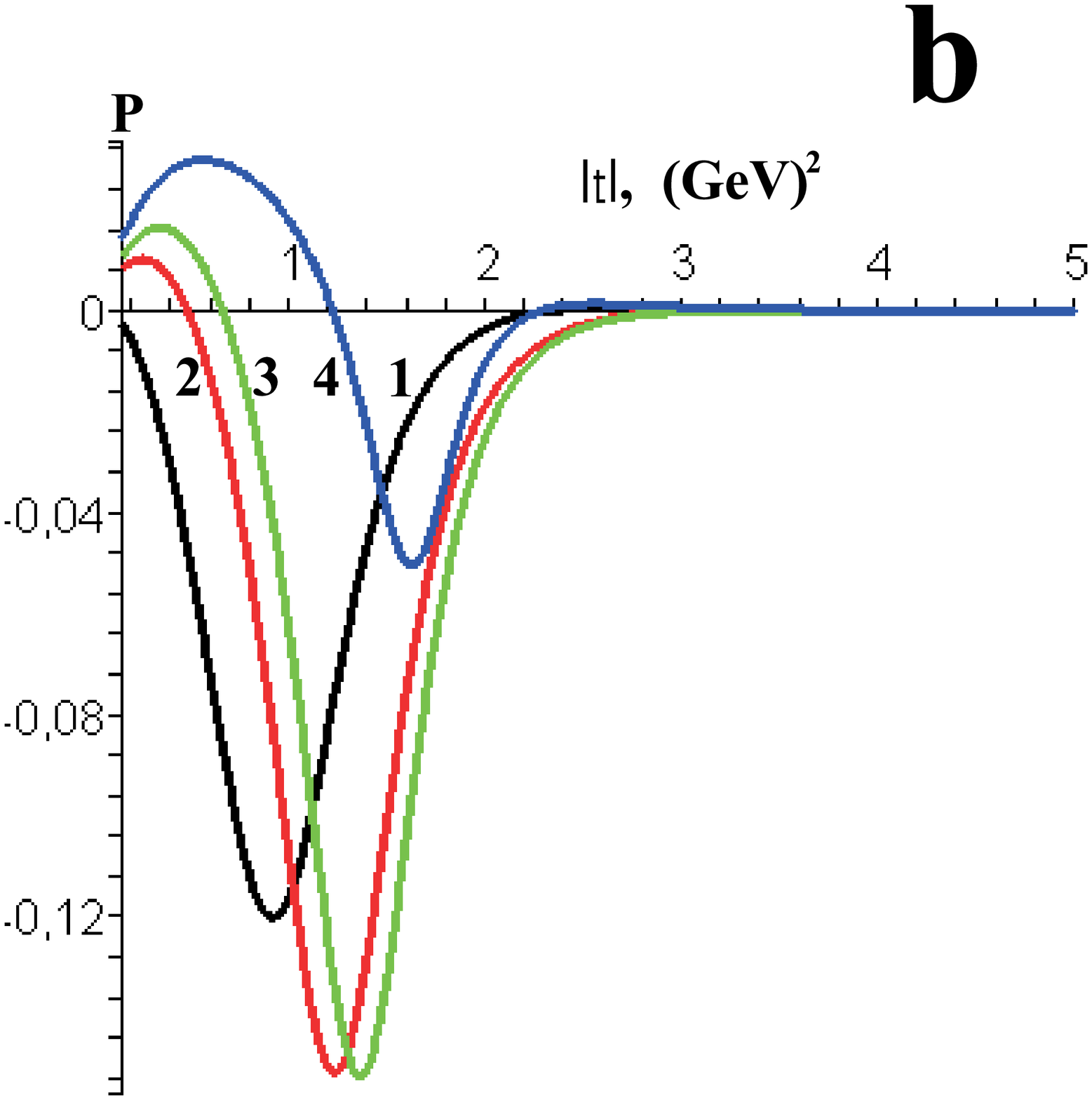}
}\label{ppbar-PAXenergies-P-version-2}}\quad
\parbox[t]{5.1cm}{%
\epsfysize=51mm \centerline{
\epsfbox{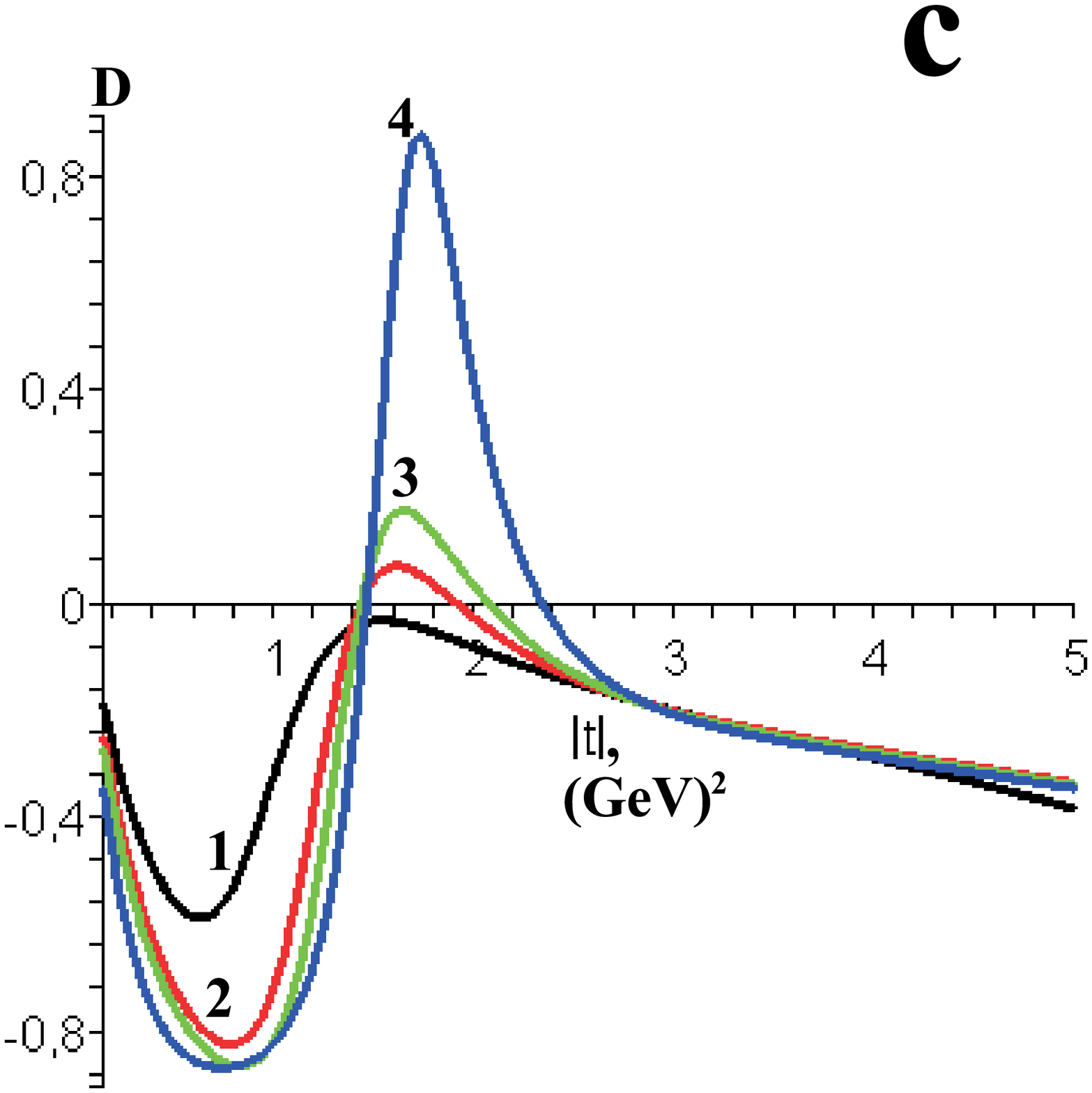}
}\label{ppbar-PAXenergies-D-version-2}}\\\\
\parbox[t]{5.1cm}{%
\epsfysize=51mm \centerline{
\epsfbox{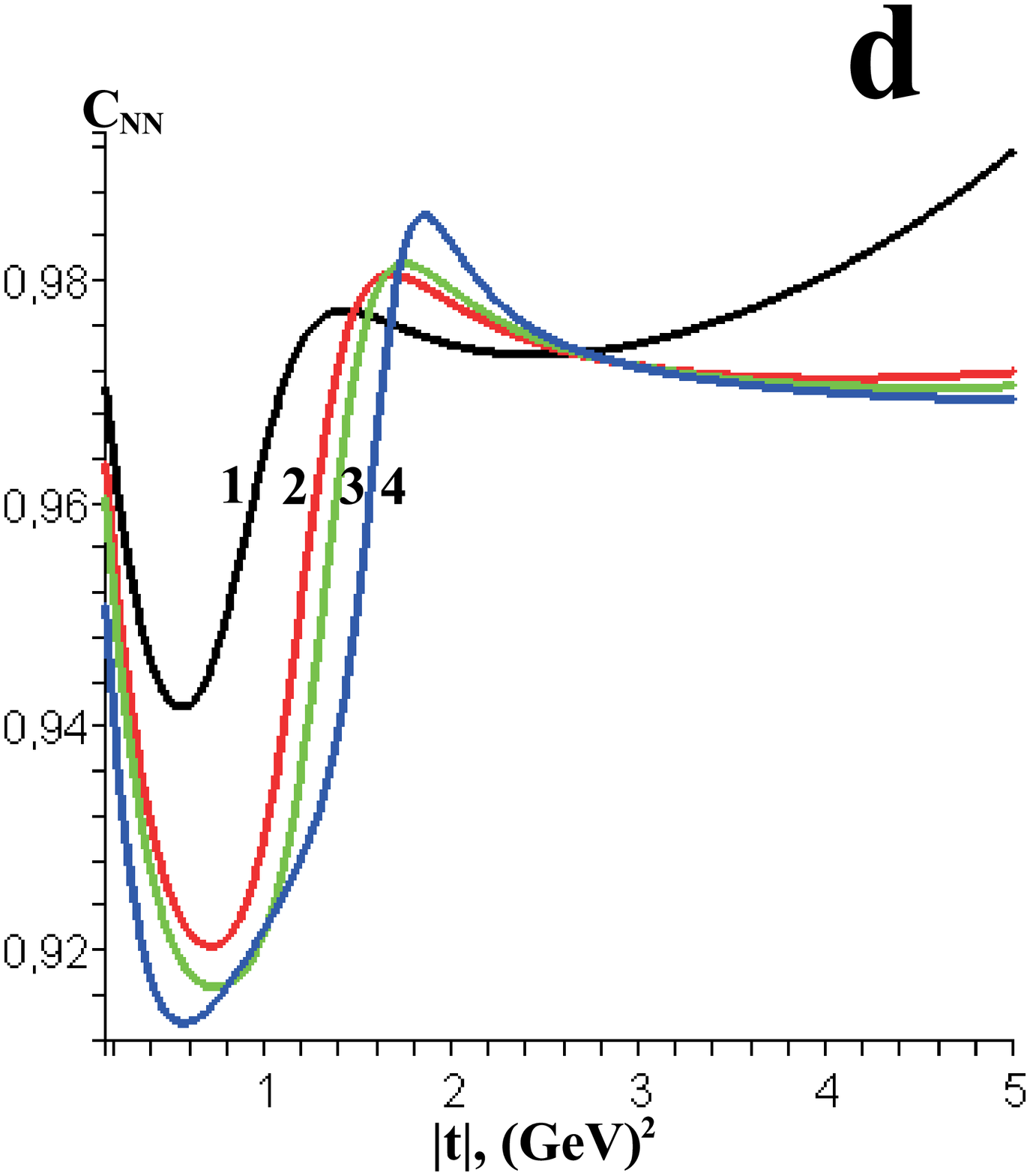}
}\label{ppbar-PAXenergies-CNN-version-2}}\quad
\parbox[t]{5.1cm}{%
\epsfysize=51mm \centerline{
\epsfbox{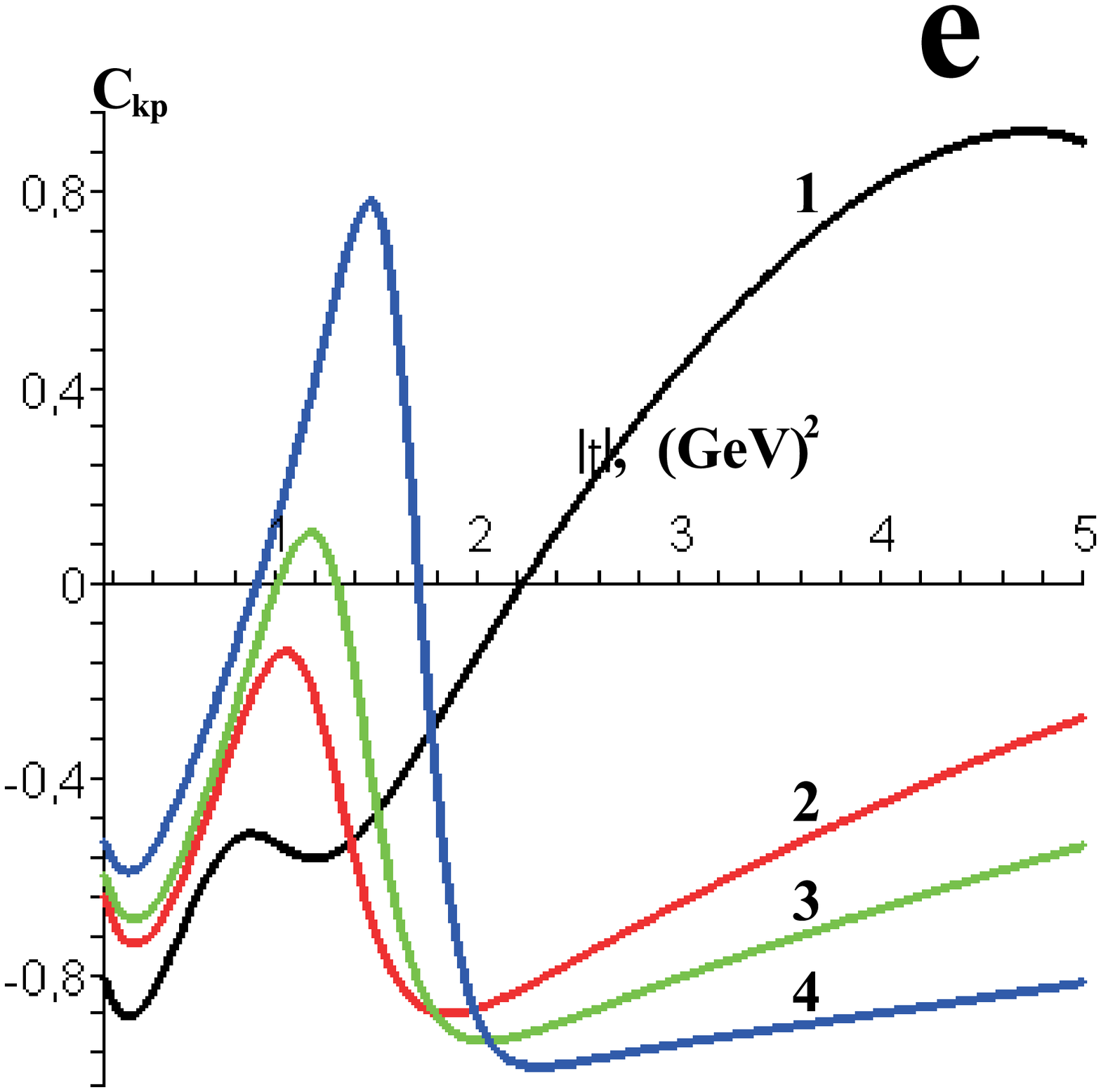}
}\label{ppbar-PAXenergies-Ckp-version-2}}\quad
\parbox[t]{5.1cm}{%
\epsfysize=51mm \centerline{
\epsfbox{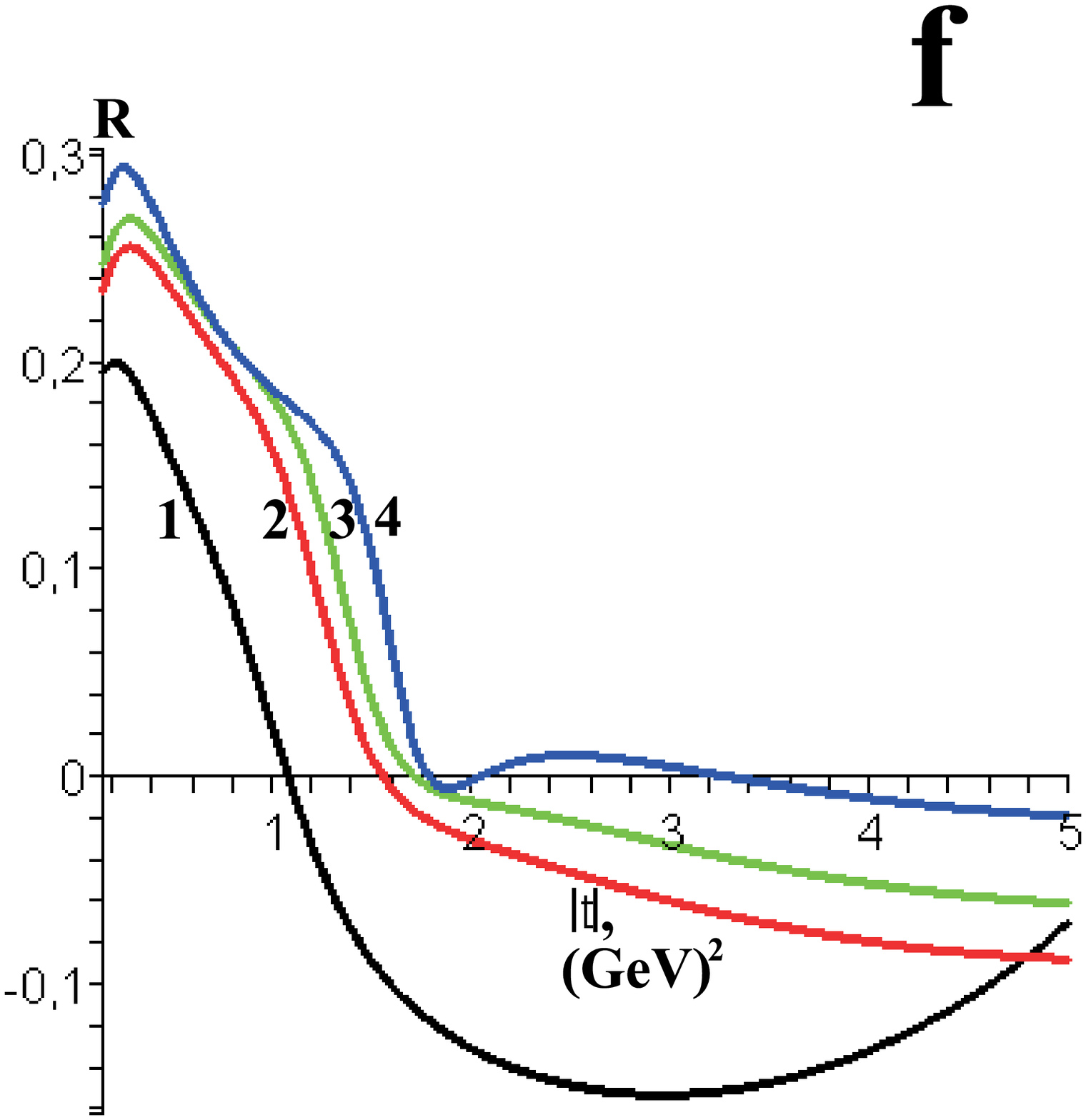}
}\label{ppbar-PAXenergies-R-version-2}}\\\\
\parbox[t]{5.1cm}{%
\epsfysize=51mm \centerline{
\epsfbox{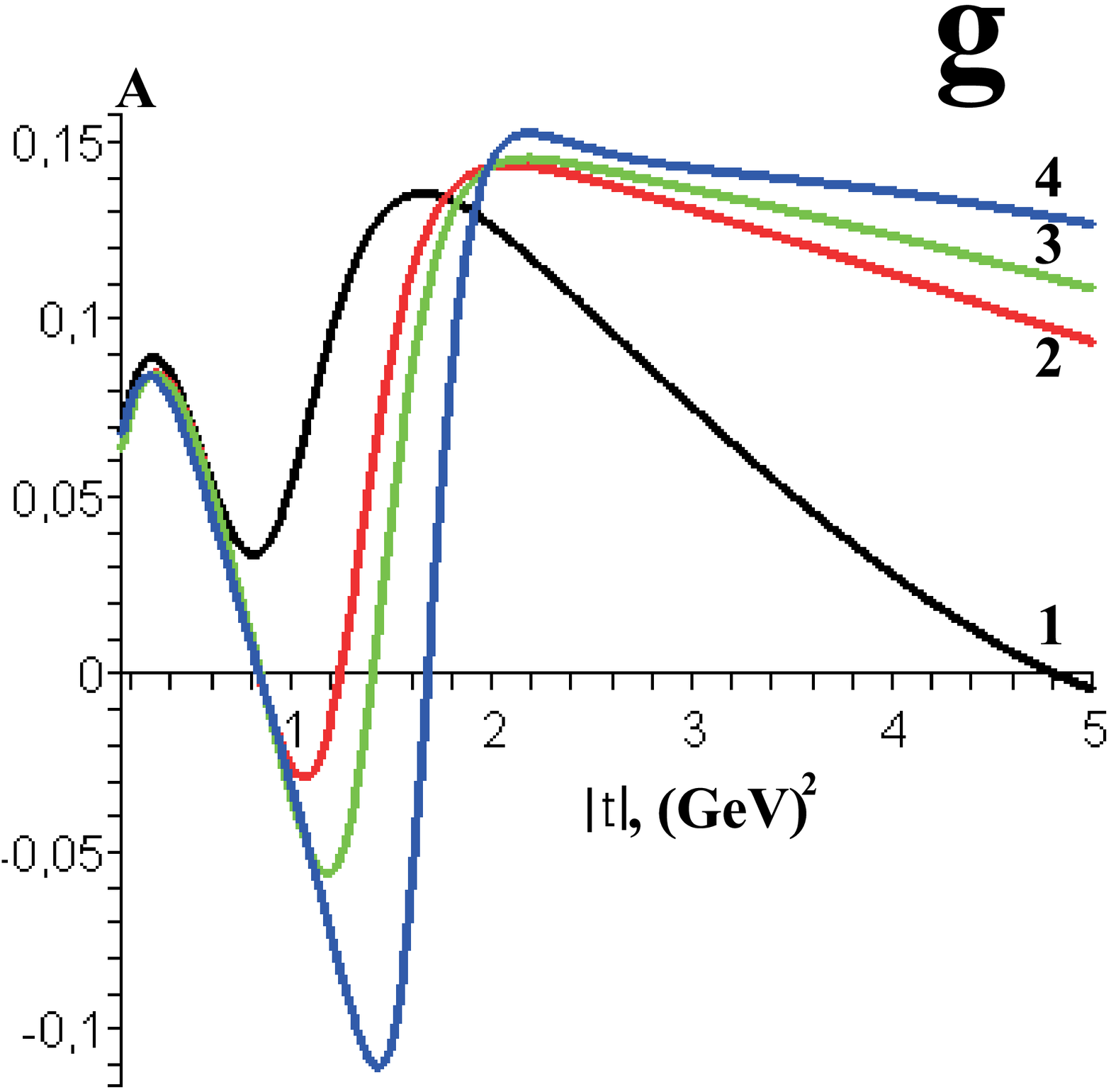}
}\label{ppbar-PAXenergies-A-version-2}}\quad
\parbox[t]{5.1cm}{%
\epsfysize=51mm \centerline{
\epsfbox{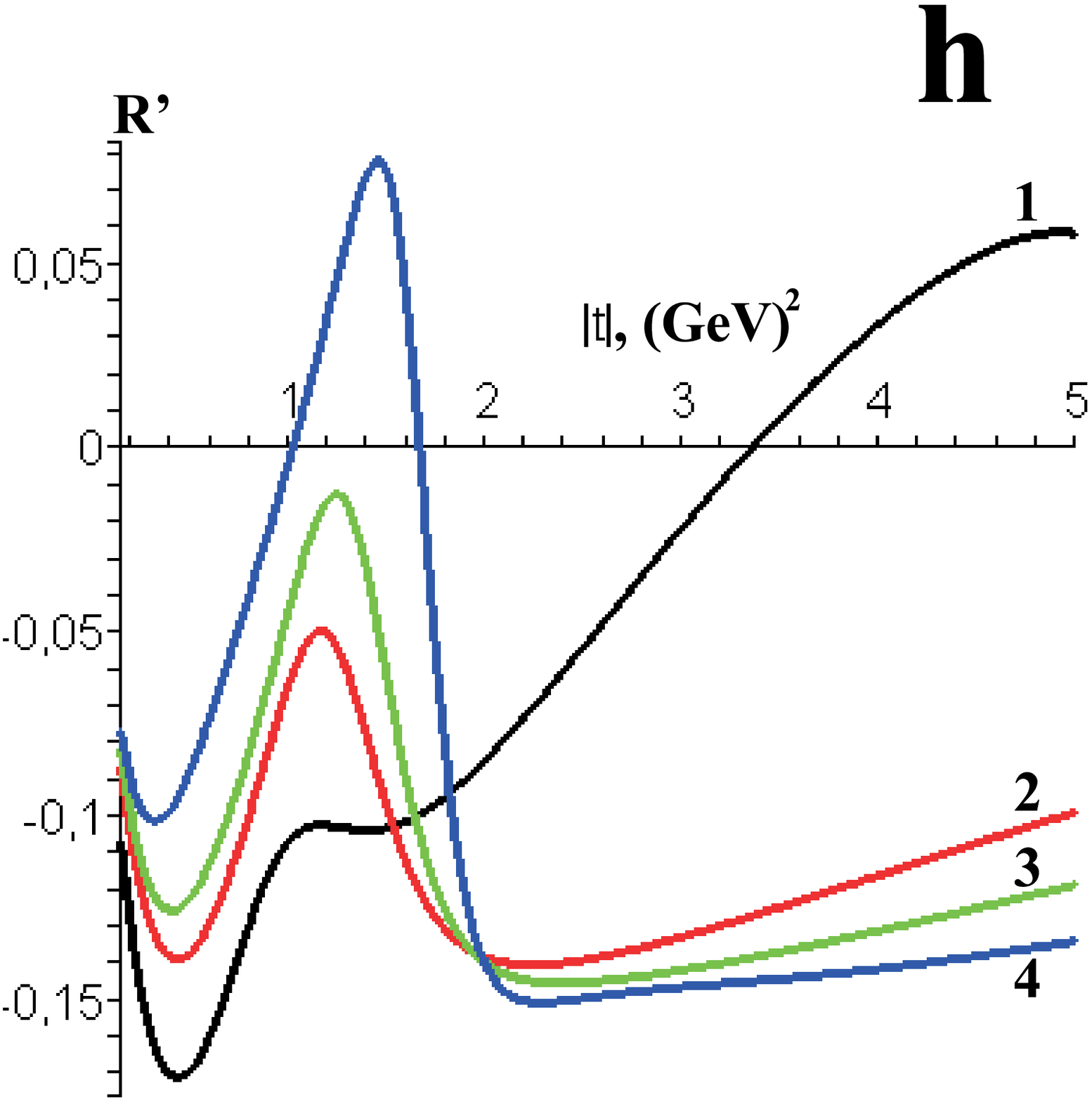}
}\label{ppbar-PAXenergies-Rprim-version-2}}\quad
\parbox[t]{5.1cm}{%
\epsfysize=51mm \centerline{
\epsfbox{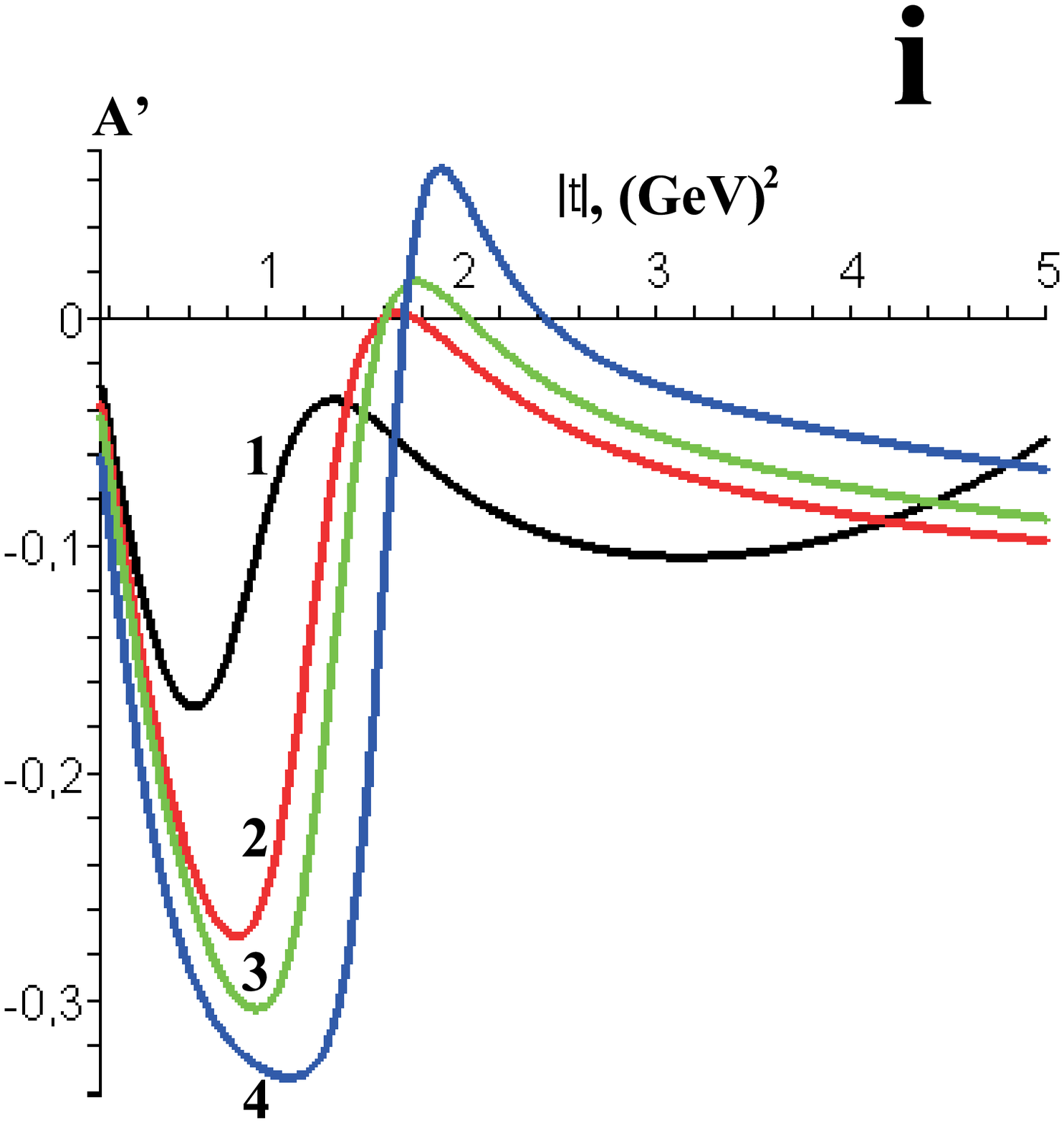}
}\label{ppbar-PAXenergies-Aprim-version-2}}
\caption{$t$-Dependence of some observable parameters for elastic
$p\bar{p}$ scattering in intermediate (PAX) energy domain: 1 -
$\sqrt{S}=3$ GeV (black), 2 - $\sqrt{S}=5$ GeV (red), 3 -
$\sqrt{S}=6.5$ GeV (green), 4 - $\sqrt{S}=14.7$ GeV (blue). The
second version of additional relations between $pp$ helicity
amplitudes is under consideration: \emph{a} - differential cross
section, \emph{b} - polarization, \emph{c} - depolarization
parameter, \emph{d} - correlation of the normal components of
polarization, \emph{e} - correlation of the transverse components
of polarization, \emph{f} - transverse polarization rotation
parameter, \emph{g} - longitudinal polarization rotation
parameter, \emph{h} - correlation of transverse-longitudinal
components of polarization, \emph{i} - correlation of
longitudinal-longitudinal components of polarization.}
\end{figure}

\end{document}